\documentclass[sigconf,nonacm]{acmart}

\usepackage{subfigure}
\usepackage{listings}
\usepackage{hyperref}
\usepackage[ruled,vlined]{algorithm2e}

\AtBeginDocument{%
  \providecommand\BibTeX{{%
    \normalfont B\kern-0.5em{\scshape i\kern-0.25em b}\kern-0.8em\TeX}}}

\setcopyright{acmcopyright}
\copyrightyear{2018}
\acmYear{2018}
\acmDOI{10.1145/1122445.1122456}

\begin{document}

\newcommand{\fftlib}{{tcFFT}} 

%%
%% The "title" command has an optional parameter,
%% allowing the author to define a "short title" to be used in page headers.
\title{\fftlib{}: Accelerating Half-Precision FFT through Tensor Cores}

%%
%% The "author" command and its associated commands are used to define
%% the authors and their affiliations.
%% Of note is the shared affiliation of the first two authors, and the
%% "authornote" and "authornotemark" commands
%% used to denote shared contribution to the research.
\author{Binrui Li}
\affiliation{%
  \institution{Shanghai Jiao Tong University}
  \city{Shanghai}
  \country{China}}
\email{rox906@sjtu.edu.cn}

\author{Shenggan Cheng}
\affiliation{%
\institution{Shanghai Jiao Tong University}
  \city{Shanghai}
  \country{China}}
\email{chengshenggan@sjtu.edu.cn}

\author{James Lin}
\affiliation{%
\institution{Shanghai Jiao Tong University}
  \city{Shanghai}
  \country{China}
}
\email{james@sjtu.edu.cn}

%%
%% By default, the full list of authors will be used in the page
%% headers. Often, this list is too long, and will overlap
%% other information printed in the page headers. This command allows
%% the author to define a more concise list
%% of authors' names for this purpose.
\renewcommand{\shortauthors}{Trovato and Tobin, et al.}

%%
%% The abstract is a short summary of the work to be presented in the
%% article.
\begin{abstract}
  Fast Fourier Transform (FFT) is an essential tool in scientific and engineering computation. The increasing demand for mixed-precision FFT has made it possible to utilize half-precision floating-point (FP16) arithmetic for faster speed and energy saving. Specializing in lower precision, NVIDIA Tensor Cores can deliver extremely high computation performance. However, the fixed computation pattern makes it hard to utilize the computing power of Tensor Cores in FFT. Therefore, we developed \fftlib{} to accelerate FFT with Tensor Cores. Our \fftlib{} supports batched 1D and 2D FFT of various sizes and it exploits a set of optimizations to achieve high performance: 1) single-element manipulation on Tensor Core fragments to support special operations needed by FFT; 2) fine-grained data arrangement design to coordinate with the GPU memory access pattern. We evaluated our \fftlib{} and the NVIDIA cuFFT in various sizes and dimensions on NVIDIA V100 and A100 GPUs. The results show that our \fftlib{} can outperform cuFFT 1.29x-3.24x and 1.10x-3.03x on the two GPUs, respectively. Our \fftlib{} has a great potential for mixed-precision scientific applications.
\end{abstract}

%%
%% The code below is generated by the tool at http://dl.acm.org/ccs.cfm.
%% Please copy and paste the code instead of the example below.
%%
  

%%
%% Keywords. The author(s) should pick words that accurately describe
%% the work being presented. Separate the keywords with commas.

%%
%% This command processes the author and affiliation and title
%% information and builds the first part of the formatted document.
\maketitle

\section{Introduction}

Fast Fourier transform (FFT) is essential in many scientific and engineering applications, including large-scale simulations \cite{cheng2020cube}, time series \cite{watson1982time}, waveform analysis \cite{biwer2019pycbc}, electronic structure calculations \cite{haynes2000parallel}, and image processing \cite{despres2017review}. Due to its wide range of applications, improving the performance of FFT is of great significance. Many efforts have been made from algorithm and hardware aspects. Lots of optimized implementations of FFT have been proposed on the CPU platform \cite{frigo1998fftw, gholami2015accfft}, the GPU platform \cite{chen2010large, nukada2012scalable} and other accelerator platforms \cite{park2013tera, liu2014memory, schwaller2017investigating}.

The demand for mixed-precision FFT is also increasing, while half precision (FP16) is gaining popularity with its faster speed and energy saving ability \cite{markidis2018nvidia}. Most of the popular FFT frameworks such as cuFFT, FFTW for ARM (Arm Performance Libraries), Vulkan FFT, include support for half precision besides single and double precision.
% Meanwhile, the demand for lower-precision or mixed-precision FFTs is gradually increasing.
And a noticeable number of scientific applications use half-precision FFT.
The gravitational wave data analysis software pyCBC \cite{biwer2019pycbc} and the cosmological large-scale structure N-body code CUBE \cite{yu2018cube, cheng2020cube} use half precision to speed up the long-length FFT calculation. Medical image restoration applications \cite{maass2011ct, despres2017review} use lower precision or mixed precision to speed up the computation of batched 2D FFT. Specializing in lower precision, NVIDIA Tensor Cores can deliver extremely high performance which makes it worthwhile to take on the challenge of designing and implementing novel FFT algorithms on them.

% In large computational applications, including image processing, speech recognition, and large scale simulations, a majority of execution time is allotted to computing the FFT. In order to improve performance of the FFT, many investigations have been made on implementing the FFT on the computationally superior Graphical Processing Unit (GPU) platform. The NVIDIA CUDA Fast Fourier Transform library (cuFFT) provides a simple interface for computing FFTs efficiently on NVIDIA GPU which is one of the most important and widely used numerical algorithms in computational physics and general signal processing.

% The recent success of machine learning (ML), or deep learning (DL) in particular, has spurred a new wave of hardware accelerators. Many specialized cores and hardware accelerators have been built to speed up the general matrix multiply (GEMM) in deep learning applications. These specialized hardware typically exploit low-precision matrix computation to achieve high performance, based on the fact that deep learning workloads involve many matrix operations and are usually robust to low-precision computation. One example is the Tensor Core on NVIDIA Volta GPUs that conduct half-precision matrix-matrix computation, achieving 8× higher throughput over the CUDA Cores. It is significant to take on the challenge of designing and engineering novel algorithms exploiting the compute power available in low precision to the application specific needs.

However, there are still following challenges to accelerate FFT with Tensor Cores which are specialized for GEMM operations. % although prior work has pointed out how to express FFT in the form of matrix multiplication.
The first challenge is how to efficiently support FFT's special operations on Tensor Cores.
Second, memory can easily become the bottleneck of FFT algorithms due to their modest arithmetic intensity and unique memory access pattern. To give full play to the computation performance of Tensor Cores, the algorithm needs to be well-designed in data arrangement.
% Second, unlike GEMM, each iteration of FFT is a small-matrix multiplication, and there is a dependency on the data between iterations. This makes the original FFT algorithm mainly limited by global memory bandwidth. To give full play to the computation performance of Tensor Core, the algorithm needs to be adjusted to alleviate the memory bottleneck to get high performance.
% Third, the algorithm needs careful design to obtain relatively consistent performance results in various sizes and dimensions FFTs.
Third, FFTs of different sizes use different kernels. Large size FFTs use more complicated kernels. Therefore, it is difficult to obtain a relatively consistent optimization effect on various FFT sizes.

% Several approaches \cite{sorna2018optimizing, cheng2018accelerating, durrani2021fft} have been proposed for using Tensor Core in computation of FFT. However, Sorna \cite{sorna2018optimizing} and Cheng \cite{cheng2018accelerating} just focused on the method of mapping FFT algorithm into GEMM operators while the performance of their implementations based on cuBlas was far inferior to cuFFT APIs. Durran \cite{durrani2021fft} implemented an optimizing 1D FFT algorithm, which outperformed cuFFT, but they only focused on the 1D FFT of limited sizes, which is just the basic simple situation. Their work has not involved many key issues that may be faced when developing a high-performance FFT library which can perform well on both 1D and 2D FFTs of various sizes. However, these are of vital importance in the main application scenarios of half-precision FFT \cite{schwaller2017investigating}.

Several approaches \cite{sorna2018optimizing, cheng2018accelerating, durrani2021fft} have been proposed for using Tensor Cores in the computation of FFT. Sorna \cite{sorna2018optimizing} and Cheng \cite{cheng2018accelerating} gived the theoretical basis and an example implementation which resorted to cuBlas to utilize Tensor Cores. But the performance of their implementation is far inferior to cuFFT. In Durran's poster \cite{durrani2021fft}, their implementation with Tensor Core WMMA APIs outperformed cuFFT, but only on the basic small size 1D FFT. They did not deal with the memory bottleneck caused by the unique memory access pattern of large size or multidimensional FFT, and there is still considerable room for improvement in their method to support FFT's special operations.
%, and they did not open their source code or give out their design in detail.

% In a word, prior work has not solved some key issues when developing a Tensor Core FFT which can perform well on both 1D and 2D FFTs of various sizes. Even though these sizes are of vital importance in the main application scenarios of half-precision FFT \cite{schwaller2017investigating}.

We designed and implemented \fftlib{}, the first FFT library on Tensor Cores which supports batched 1D and 2D FFT in a wide range of sizes with high performance, and it is open-source at \url{https://github.com/given_in_the_official_version}. It can outperform cuFFT in common half-precision FFT applied scenarios \cite{biwer2019pycbc, yu2018cube, cheng2020cube, maass2011ct, despres2017review} and uses the similar interface to cuFFT. We have overcome the key challenges in implementing such a universal size supported FFT library with two major novel techniques.
(1) First, FFT's special operations, complex matrix accesses and element-wise multiplications, are not natively support by Tensor Cores. This reduces the benefits of using Tensor Cores. To perform these operations more efficiently, we proposed a method to operate Tensor Core fragments with single element granularity according to the map of matrix elements into each thread's registers.
(2) Second, merging processes of large size or 2D FFT require strided memory accesses, and uncoalesced strided global memory accesses are quite inefficient on GPU. Besides using shared memory to reduce the number of global memory accesses, we redesigned the data layout in memory and implemented a continuous memory access pattern with variable size. This pattern can ensure that there are sufficient continuous elements when accessing the data. 

In summary, this paper makes the following contributions.
\begin{itemize}
  \item We developed \fftlib{} to accelerate FFT on Tensor Cores with high performance, which supports batched 1D and 2D FFT of various sizes. (Sec \ref{sec:design}).
  \item We propose two major novel techniques to imporve its performance: a) a single-element manipulation on Tensor Core fragments to efficiently support special operations needed by FFT (Sec \ref{sec:opt_TC}); b) a special design for data arrangement and a continuous memory access pattern with variable size to alleviate the memory bottleneck (Sec \ref{sec:opt_mem});
  \item We evaluate \fftlib{} on NVIDIA Volta and Ampere GPUs. On V100, it achieves 1.90x speedup on average on 1D FFTs and 1.29x-3.24x speedup on 2D FFTs over half-precision kernels on CUDA cores from cuFFT. On A100, it achieves 1.24x on average and 1.10x-3.03x, respectively (Sec \ref{sec:results}).
\end{itemize}

\section{Background}
% In this section, we first gives the mathematical deduction of the most common FFT algorithm, Cooley-Tukey Algorithm, this is the theoretical basis for expressing FFT in matrix form. Then, we point out the important features of Tensor Cores very briefly.

\subsection{FFT in Matrix Form}
\label{sec:bg_FFT}

Fast Fourier transform is an efficient algorithm to compute the discrete Fourier transform(DFT) of a sequence. The DFT converts an N-point sequence into a same-length sequence, according to the following equation, where $x$ denotes the original sequence and $X$ denotes its DFT:
\begin{equation}
  X[k]=\sum_{n=0}^{N-1}x[n]\times e^{-j2\pi nk/N}
  \label{equation:dft_def}
\end{equation}
This process can be viewed as a matrix-vector multiplication between the DFT matrix and the original sequence. The computational complexity of this process is $O(N^{2})$. The DFT is useful in many fields, but computing it directly from the definition is often too slow to be practical. One of the most widely used FFT algorithm, Cooley-Tukey FFT algorithm, reduce the computational complexity to $O(Nlog(N))$. This algorithm can calculate the DFT of an N-point sequence from $N_1$ DFTs of its $N/N_1$-point subsequences according to equation \ref{equation:dft_s2} which can be used recursively until the size of the subsequences is 1. The basic flow of the algorithm has three steps:

% The algorithm is efficient because of the following facts:

% Let $W_{N}^{nk}=e^{-j2\pi nk/N}$, equation \ref{equation:dft_def} can be rewritten into equation \ref{equation:dft_s1} where the sequence $x$ is rearranged into $N_1$ parts. Then let $x_m$ denotes the m-th $(N_2-1)$-point sampling subsequence of $x$, where $x_m[n]=x[nN_1+m]$, for all $0<=n<=N_2-1$, and let $X_m$ dentoes its DFT, $\sum_{n=0}^{N_{2}-1}x[nN_{1}+m]\times W_{N_{2}}^{kn}$ can be rewritten to $X_{m}[k]$, when $0<=k<=N_2-1$. Moreover, using the periodicity of $W_{N_2}^{kn}$, equation \ref{equation:dft_s1} can be rewritten to equation \ref{equation:dft_s2} for $0<=k<=N-1$. This equation expresses the DFT of an N-point sequence in terms of $N_1$ DFTs of $N/N_1$-point sequences, which can be used recursively untill the size of the sequence is 1. The basic flow of the algorithm is as follows:

\begin{enumerate}
  \item Divide the original N-point sequence into $N_{1}$ subsequences of length $N_{2}$ which equals $N/N_1$.
  \item Use Cooley-Tukey FFT algorithm on each $N_2$-point subsequence recursively.
  \item Combine $N_1$ subsequences to get the DFT of the original N-point sequence, according to equation \ref{equation:dft_s2}.
\end{enumerate}

The above radix-$N_1$ Cooley-Tukey algorithm reduces the computational complexity to $O(N_1N\log_{N_1}{N})$. $\log_{N_{1}}{N}$ is the number of the major recurrences and ${N_{1}}N$ is the number of operations in each iteration, where $N_{1}$ usually takes the value 2 or 4 to simplify the calculation in traditional implementations of FFT.

% This process, reducing the computational complexity from $O(N^2$ $)$ to $O(N_1N\log_{N_1}{N})$, is the core of radix-$N_1$ Cooley-Tukey FFT algorithm. $\log_{N_{1}}{N}$ decides the number of the major recurrences and ${N_{1}}N$ decides the number of operations of each iteration, where $N_{1}$ usually takes 2 or 4 to minimize the number of total calculations in traditional implementations of FFT.

% \begin{equation}
%   \label{equation:dft_s1}
%   \begin{aligned}
%     X[k]=&\sum_{n=0}^{N-1}x[n]\times W_{N}^{kn}\\
%     =&\sum_{m=0}^{N_{1}-1}\sum_{n=0}^{N_{2}-1}x[nN_{1}+m]\times W_{N}^{k(nN_{1}+m)}\\
%     =&\sum_{m=0}^{N_{1}-1}W_{N}^{mk} \sum_{n=0}^{N_{2}-1}x[nN_{1}+m]\times W_{N_{2}}^{kn}
%   \end{aligned}
% \end{equation}

\begin{equation}
  \label{equation:dft_s2}
  \begin{aligned}
    X[k]=&\sum_{m=0}^{N_{1}-1}W_{N}^{mk}X_{m}[k \  mod \  N_2]
  \end{aligned}
\end{equation}

% \subsection{Re-express FFT with Matrix Form}
We call step (3) of the original algorithm a merging process for the rest of the paper. It is the core of FFT. The complete FFT algorithm consists of multiple merging processes. The merging process can be expressed in the form of matrix as equation \ref{equation:dft_mm}. It is rewritten from equation \ref{equation:dft_s2} according to the periodicity of $W_N^{mk}$.
% \begin{equation}
%   \label{equation:dft_s3}
%   \begin{aligned}
%     X[k+lN_{2}]=&\sum_{m=0}^{N_{1}-1} W_{N}^{m(k+lN_{2})} X_{m}[k]\\
%     =&\sum_{m=0}^{N_{1}-1} W_{N_{1}}^{ml} ( W_{N}^{mk} X_{m}[k] )
%   \end{aligned}
% \end{equation}
\begin{equation}
  \label{equation:dft_mm}
  X_{out} = F_{N_1} \cdot (T_{N_1N_2} \odot X_{in})\\
\end{equation}
where, $\cdot$ denotes matrix product, $\odot$ denotes element-wise product. $X_{out}$ represents the matrix form of the output N-point DFT of the original sequence. The matrix is $N_1 \times N_2$ and is shown as follows:
\begin{small}
  \begin{equation*}
    X_{out}=
    \begin{bmatrix}
      X[0] & X[1] & \cdots & X[N_2-1] \\
      X[N_2] & X[N_2+1] & \cdots & X[2N_2-1] \\
      \vdots & \vdots &  & \vdots\\
      X[(N_1-1)N_2] & X[(N_1-1)N_2+1] & \cdots & X[N_1N_2-1] \\
    \end{bmatrix}
  \end{equation*}
\end{small}
$F_{N_1}$ denotes the $N_1 \times N_1$ radix-$N_1$ DFT matrix. It can fit well in the Tensor Cores when $N_1$ is 16. $T_{N_1N_2}$ is an $N_1 \times N_2$ twiddle factor matrix, it has the same size as $X_{out}$ and $X_{in}$. $F_{N_1}$ and $T_{N_1N_2}$ are shown as follows:
\begin{equation*}  
  F_{N_1}=
  \begin{bmatrix}
    W_{N_1}^0 & W_{N_1}^0 & \cdots & W_{N_1}^0\\
    W_{N_1}^0 & W_{N_1}^1 & \cdots & W_{N_1}^{N_1-1}\\
    \vdots & \vdots &  & \vdots\\
    W_{N_1}^0 & W_{N_1}^{N_1-1} & \cdots & W_{N_1}^{(N_1-1)(N_1-1)}\\
  \end{bmatrix}
\end{equation*}
\begin{equation*}
  T_{N_1N_2}=
  \begin{bmatrix}
      W_N^{0} & W_N^{0} & \cdots & W_N^{0} \\
      W_N^{0} & W_N^{1} & \cdots & W_N^{N_2-1} \\
      \vdots & \vdots &  & \vdots\\
      W_N^{0} & W_N^{N_1-1} & \cdots & W_N^{(N_1-1)(N_2-1)} \\
    \end{bmatrix}
\end{equation*}
$X_{in}$ denotes the input $N_1$ $N_2$-point DFT sequences as follows:
\begin{equation*}
  X_{in}=
  \begin{bmatrix}
    X_0[0] & X_0[1] & \cdots & X_0[N_2-1] \\
    X_1[0] & X_1[1] & \cdots & X_1[N_2-1] \\
    \vdots & \vdots &  & \vdots\\
    X_{N_1-1}[0] & X_{N_1-1}[1] & \cdots & X_{N_1-1}[N_2-1] \\
  \end{bmatrix}
\end{equation*}
, where $X_m$ dentoes the DFT of the m-th $(N_2-1)$-point sampling subsequence of original $x$. $x_m[n]=x[nN_1+m]$, for all $0 \leq n \leq N_2-1$. $X_m$ is calculated in the previous iteration.

The above content describes how the FFTs of $N_1$ short sequences are converted into the FFT of a long sequence through matrix multiplication and element-wise multiplication in a merging process. Since the FFT of a sequence of length one is itself, the FFT of the original sequence can be obtained through this process of $\log_{N_1}{N}$ times.

\subsection{Tensor Cores}

Tensor Cores are special computing units introduced from NVIDIA Volta microarchitectures which perform mixed-precision matrix-multiply-and-accumulate operation. Different from CUDA Cores, Tensor Cores provide a compute primitive of matrix-matrix rather than scalar-scalar. For example, Tesla V100 contains 640 Tensor Cores. Each Tensor Core can perform the operation $D = A \cdot B + C$ in one GPU clock cycle, where all matrices are $4 \times 4$ matrices. The precision of matrices $A$ and $B$ are FP16, while The precision of matrices $C$ and $D$ may be FP16 or FP32.

% \begin{figure}[h]
%   \centering
%   \includegraphics[width=\linewidth]{pics/tensorcore}
%   \caption{Mixed Precision $4 \times 4$ Matrix Multiplication of Tensor Cores.}
%   \Description[Tensor Core]{Mixed Precision $4 \times 4$ Matrix Multiplication of Tensor Cores.}
%   \label{fig:tensorcore}
% \end{figure}

Tesla V100 and A100 deliver groundbreaking performance with half-precision or mixed-precision matrix multiply through Tensor Cores, which is shown in Table \ref{table:tc_flops}. Tensor Cores have their own local memory consisting of fragments. Matrices are loaded and stored in fragments, allowing data sharing across registers \cite{cudatensor}. Existing work reveals that fragments will use register memory from the hardware perspective \cite{jia2018dissecting, jia2019dissecting, yan2020demystifying}.

% From the perspective of hardware implementation, existing work reveals that Fragment is implemented as a register 

\begin{table}[]
  \centering
  \caption{Performance of Tensor Cores on V100 and A100.}
  \label{table:tc_flops}
  \begin{tabular}{ccc}
  \hline
  Performance            & Tesla V100     & Tesla A100   \\ \hline
  Peak FP64         & 7.8 teraFLOPS  & 9.7 teraFLOPS  \\
  Peak FP32         & 15.7 teraFLOP  & 19.5 teraFLOPS \\
  FP16 Tensor Core   & 125 teraFLOPS  & 312 teraFLOPS \\ \hline
  \end{tabular}
\end{table}

\section{\fftlib{}}
\label{sec:design}
% We first change the traditional FFT algorithm in section \ref{sec:bg_FFT} into matrix form which can be ported on Tensor Cores. It is the core of \fftlib{}. Then, we design \fftlib{} library which supports batched and 2D FFT and more sizes.

\subsection{Architecture}
\begin{figure}[tb]
  \centering
  \includegraphics[width=1\linewidth]{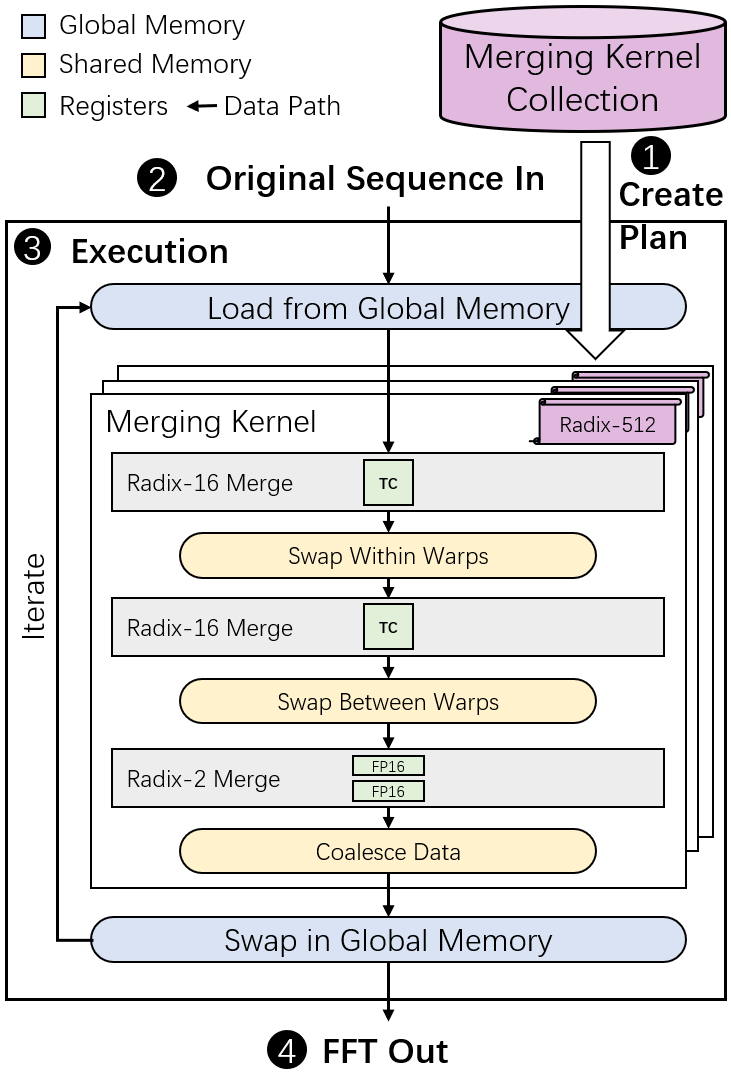}
  \vspace{-1\baselineskip}
  \caption{The complete execution prcocess of \fftlib: First, create a \textit{plan} based on the dimensions and the size of the input data. This plan selects an optimal set of \textit{merging kernels} from the pre-implemented \textit{merging kernel collection}. Then, the \textit{execution function} is called with the plan and the original data as input. In the \textit{execution function}, the previously determined merging kernels are executed in turn in multiple iterations and multiple types of memory in GPU is fully utilized to increase the reuse of data.}
  \label{fig:functions}
\end{figure}

% After obtaining equation \ref{equation:dft_mm}, we begin to design \fftlib{}.
Modeled after FFTW and cuFFT, \fftlib{} uses a simple configuration mechanism called a \textit{plan}. A \textit{plan} chooses a series of optimal \textit{radix-X merging kernels}. Then, when the \textit{execution function} is called, actual transform takes place following the plan.
Figure \ref{fig:functions} shows the complete process of performing an FFT. First, a function is called to create a \textit{plan} based on the dimension and the size of the input data. This plan selects an optimal set of \textit{merging kernels} from the pre-implemented \textit{merging kernel collection}. Then, the \textit{execution function} is called with the plan and the original data as input. In the \textit{execution function}, the previously determined merging kernels are executed in turn in multiple iterations. Meantime, multiple types of memory in GPU is fully utilized to increase the reuse of data. After all iterations are executed, the FFT of the original sequence is obtained.

By decomposing the FFT process into a series of merging kernels, we improved the code reusability and also greatly reduced the workload of subsequent performance-related optimization work.

\medskip

\noindent\textbf{Support FFTs of all power-of-two sizes.} We have developed a series of merging kernels of different radices to support FFTs of all power-of-two sizes. The radices of these kernels cover all powers of 2 from 16 to 8192 and larger size FFTs can be realized by combining these basic kernels. Tensor Cores only provide computing power for computing $16 \times 16 \times 16$ GEMM, which can be used in power-of-16 radices. To implement more radices, smaller radices should be introduced. We introduced radix 2 and radix 4, for their DFT matrices only have 0, 1 and -1, and have high computational efficiency. They are computed using FP16 CUDA Cores and account for a small proportion in the total calculation time. 

By implementing a collection of merging kernels of a lot of sizes, we enable \fftlib{} to support FFTs of all power-of-two lengths.
% Among these basic kernels, 256, 512 and 1024 are the three most commonly used radices, for 

% When only using the radix-16 decomposition method, N must be a power of 16. In order to support sequences of other sizes, smaller radix should be introduced. We introduced radix-2 and radix-4 decimation methods. These two radices use simple FFT matrices with only 0, 1 and -1, and has high computational efficiency. These radices are computed using the normal fp16 units for they are too small to make good use of Tensor Cores. These two bases account for a small proportion in the calculation process. By combining radix-2, radix-4 and radix-16 decimation methods, we enable the \fftlib{} algorithm to support sequences of all power-of-two length.

\medskip

\noindent\textbf{Support batched FFT and 2D FFT.} Batched FFT and 2D FFT are of vital importance in application scenarios. 2D FFT performs FFT on each dimension of 2D sequences in turn. It can be implemented by strided batched FFT. Take a 2D FFT on an $N_1 \times N_2$ row-major matrix as an example, an $N_1$ batch of $N_2$-point FFT is applied on the $N_1$ rows and then an $N_2$ batch of strided $N_1$-point FFT is applied on the $N_1$ columns. The order of these two processes can be changed but strided batched FFT is essential.

% Batched FFT is used to transform many sequences in one time. Multi-dimensional FFT is used to transform input data of two or more dimensions. It performs FFT on each dimension of multi-dimensional data in turn. Multi-dimensional FFT contains batched FFT. Take a 2D FFT on an $N_1 \times N_2$ matrix as an example, an $N_1$ batch of $N_2$-point FFT is applied on the $N_1$ rows and then an $N_2$ batch of $N_1$-point FFT is applied on the $N_1$ columns. It should be noted that the order of processing here can be changed and it will not affect the final result.

% Batched FFT and multi-dimensional FFT can be implemented by calling a series of \textit{radix-X merging kernels} above with proper parameters, \textit{n\_size} and \textit{istride}.

We have implemented merging kernels that support batch and strided data. Batched FFT and 2D FFT can be implemented by calling them with proper parameters. \fftlib{} has the following plan functions for batched 1D and 2D FFTs:
\begin{itemize}
  \item \textit{tcfftPlan1D (tcfftHandle *plan, int nx, int batch)}: This function is used to create a configuration to execute FFT on a batch of 1D sequences of equal length. \textit{nx} gives the length and \textit{batch} is the number of sequences.
  \item \textit{tcfftPlan2D (tcfftHandle *plan, int nx, int ny, int batch)}: This function is used for FFT on batched 2D sequences. \textit{nx} gives the size of the first dimension and \textit{ny} is the size of the second dimension. The data are stroed in row-major, which means that the second dimension continues in memory.
  %  \textit{batch} is the number of transforms.
\end{itemize}
These functions can create plans for batched 1D FFT and 2D FFT. They pre-selected a series of optimal merging kernels of different radices for the special size.

\subsection{Merging Kernels}

\begin{algorithm}
  \caption{Radix-512 merging kernel}
  \label{alg:merge}
  \SetAlgoLined
  \LinesNumbered
  \KwIn{$N_1 \times 512$ FFTs of size $N_2$}
  \KwOut{$N_1$ FFTs of size $512N_2$}
  $F_{16\times16} \leftarrow$ Load Radix-16 DFT Matrix\;
  $Tw_{16\times N_2} \leftarrow$ Prepare Twiddle Factors while read Input\;
  Consider input data as $32N_1$ $In_{16\times N_2}$ matrices\;

  \ForEach{$In_{16 \times N_2 matrix}$}{
    \ForEach{$InFrag_{16\times16}$ in $In_{16 \times N_2}$}{
      \tcc{Parallel by Warps}
      $TwFrag_{16\times16} \leftarrow$ load a fragment from $Tw_{16\times N_2}$\;
      $InFrag_{16\times16} \leftarrow InFrag_{16\times16} \odot TwFrag_{16\times16}$ \tcp*{on FP16 units}
      $OutFrag_{16\times16} \leftarrow F_{16\times16} \cdot InFrag_{16\times16}$ \tcp*{on Tensor Cores}
      Store $OutFrag_{16\times16}$ to intermediate data\;
    }
  }

  $Tw_{16\times 16N_2} \leftarrow$ Prepare Twiddle Factors\;
  Consider intermediate data as $2N_1$ $In_{16\times 16N_2}$ matrices\;

  \ForEach{$In_{16 \times 16N_2 matrix}$}{
    \ForEach{$InFrag_{16\times16}$ in $In_{16 \times 16N_2}$}{
      \tcc{Parallel by Warps}
      $TwFrag_{16\times16} \leftarrow$ load a fragment from $Tw_{16\times 16N_2}$\;
      $InFrag_{16\times16} \leftarrow InFrag_{16\times16} \odot TwFrag_{16\times16}$ \tcp*{on FP16 units}
      $OutFrag_{16\times16} \leftarrow F_{16\times16} \cdot InFrag_{16\times16}$ \tcp*{on Tensor Cores}
      Store $OutFrag_{16\times16}$ to intermediate data\;
    }
  }

  $F_{2\times2} \leftarrow$ Load Radix-2 DFT Matrix\;
  $Tw_{2\times 256N_2} \leftarrow$ Prepare Twiddle Factors\;
  Consider intermediate data as $N_1$ $In_{2\times 256N_2}$ matrices\;

  \ForEach{$In_{2 \times 256N_2 matrix}$}{
    \ForEach{$InFrag_{2\times1}$ in $In_{2 \times 256N_2}$}{
      \tcc{Parallel by Threads}
      $TwFrag_{2\times1} \leftarrow$ load a fragment from $Tw_{2\times 256N_2}$\;
      $InFrag_{2\times1} \leftarrow InFrag_{2\times1} \odot TwFrag_{2\times1}$ \tcp*{on FP16 units}
      $OutFrag_{2\times1} \leftarrow F_{2\times2} \cdot InFrag_{2\times1}$ \tcp*{on Tensor Cores}
      Store $OutFrag_{2\times1}$ to output data\;
    }
  }
\end{algorithm}

Algorithm \ref{alg:merge} shows the radix-512 batched merging kernel. It takes $N_1$ groups of data as input. There are 512 FFT sequences of size $N_2$ in each group, and they will be merged into one FFT sequence of size $512N_2$. The merging kernel contains three sub-merging processes. Line 1-11 and line 12-21 are two radix-16 sub-merging processes accelerated by Tensor Cores and line 22-32 is a radix-2 sub-merging process. Radix-16 sub-merging kernel is the base of \fftlib{}.

\medskip

\noindent\textbf{Radix-16 sub-merging kernel.} \textit{Radix-16 merging kernel} is the base of \fftlib{}, because a $16 \times 16$ matrix can fill a Tensor Core fragment, bringing the highest computational efficiency. We treat the $X_{out}$ matrix and the $X_{in}$ matrix as multiple $16 \times 16$ matrices, which can be calculated in parallel. A \textit{radix-16 sub-merging kernel} can execute a radix-16 merging process described by the equation $X_{out} = F_{N_1} \cdot (T_{N_1N_2} \odot X_{in})$ from section \ref{sec:bg_FFT}. Radix-16 merging process combines FFTs of 16 sequences into an FFT of a sequence of 16 times the length. Strided data layout is also supported, and we will optimize the efficiency of memory accesses in section \ref{sec:opt_mem}.

As shown in line 1-11 in algorithm \ref{alg:merge}, the sub-merging process firstly loads the radix-16 DFT matrix and calculates twiddle factors while reading input. These matrices are divided into $16 \times 16$ fragments and distributed to the GPU warps for parallelization. Then wraps execute matrix-multiplication with Tensor Cores and element-wise multiplication with FP16 CUDA Cores on these fragments. After that, all fragments are put together, and $N_1 \times 512$ FFT sequences of size $N_2$ are merged into $N_1 \times 32$ FFT sequences of size $16N_2$. Throughout the sub-merging process, intermediate results are stored in the Tensor Core fragments, we will discuss how to manipulate them efficiently in section \ref{sec:opt_TC}.

% We design \textit{radix-16 sub-merging kernel} which can both support batched 1D FFT and multi-dimensional FFT. Function \textit{radix\_16\_mer ge(half2 *data, int istride, int n\_size, int n\_sequence)} can carry out an in-place radix-16 merge on the input data.

% As shown in picture \ref{fig:radix4-merge}, the input data consists of FFTs of \textit{n\_sequence} sequences of \textit{n\_size} length. The input data will be treated as a matrix, \textit{n\_size} actually gives the length of the rows of the matrix and \textit{istride} gives the distance between the first elements of adjacent rows in the matrix. \textit{n\_size} and \textit{istride} are same when doing FFT on 1D sequence or batched 1D sequences, but they are different when doing FFT on multi-dimensional data. For example, when doing FFT along columns of a row-major matrix, the initial \text{n\_size} is one but the initial \textit{istride} is the length of a row in the original matrix.

% \begin{figure*}[h]
%   \centering
%   \includegraphics[width=\linewidth]{pics/radix4-merge}
%   \caption{A radix-16 merging process: the input data consists of FFTs of \textit{n\_sequence} sequences of \textit{n\_size} length. They are treated as a matrix, and \textit{istride} gives the distance between the first elements of adjacent rows in the matrix. After this merging process, \textit{n\_sequence} sequences are merged into \textit{n\_sequence / 16} sequences. The length of each sequences and the \textit{istride} becomes 16 times the original.}
%   \label{fig:radix4-merge}
% \end{figure*}

\medskip

\noindent\textbf{Combine multiple mergings.} The $\log_{}{N}$ merging processes of the FFT algorithm requires $\log_{}{N}$ times of memory accesses, and the arithmetic intensity of an original merging process is not large enough. To reduce the times of global memory accesses and increase the arithmetic intensity, a complete merging kernel consists of multiple sub-merging processes and uses shared memory to exchange data in the middle.

The range of data exchange taking place in the whole FFT process has different scales. Take radix-512 merging kernel as an example, after the first radix-16 merging, each warp can exchange data internally through shared memory, for it holds all the elements needed. This can be done without synchronization. After the second radix-16 merging, data are exchanged between warps also through shared memory, but a block-range synchronization is needed. After the second radix-2 merging, data exchanges between blocks are needed, which can only be done through global memory.

% In the middle of a large radix calculation, \fftlib{} perform data exchanges in two levels.
% We have implemented radix-256, radix-512, radix-1024 and other commonly used large radix merging kernels by combine radix-16, radix-4 and radix-2 sub-merging kernels.

% This optimization helps to reduce the times of global memory accessing to half, with only unavoidables synchronization introduced.

We implemented merging kernels so that global memory accesses are performed only when necessary and only unavoidable synchronizations are introduced. This implementation achieves the purpose of reducing bandwidth requirements and increasing arithmetic intensity.

\section{Performance Optimization}
% In order to take full advantage of the performance improvement brought by Tensor Cores, we have made the following optimizations on the basic algorithm.

\subsection{Optimizations to Tensor Cores}
\label{sec:opt_TC}

FFT requires complex-matrix access and element-wise multiplication operations. However, they can not be performed efficiently when the data is stored as a Tensor Core fragment due to the limitations of Tensor Core APIs.

% Tensor Cores don't natively support complex-matrix multiplication. We decompose complex-matrix multiplication $C_{complex}=A_{complex} \cdot B_{complex}$ into real-matrix multiplication as follows:
% \begin{equation}
%   \begin{aligned}
%     C_{real} \leftarrow & A_{real} \cdot B_{real} - A_{imag} \cdot B_{imag}\\
%     C_{imag} \leftarrow & A_{real} \cdot B_{imag} + A_{imag} \cdot B_{real}
%   \end{aligned}
% \end{equation}

% However, there are difficulties when trying to acess complex data efficiently and to do element-size multiplication efficiently, due to the limitations on Tensor Core WMMA APIs.

\medskip

\noindent\textbf{The limitations of Tensor Core APIs.} NVIDIA provides Warp Matrix Multiply Accumulate (WMMA) APIs for leveraging Tensor Cores to accelerate matrix problems of the form $D=A \cdot B+C$. There are four functions as follows \cite{cudac++programmingguide}:
\begin{itemize}
  \item \textit{load\_matrix\_sync}: All warp lanes loads a matrix fragment from memory, synchronously.
  \item \textit{store\_matrix\_sync}: All warp lanes stores a matrix fragment to memory, synchronously.
  \item \textit{fill\_fragment}: Fill a matrix fragment with a constant value.
  \item \textit{mma\_sync}: All warp lanes perform the warp-synchronous matrix multiply-accumulate operation $D=A \cdot B+C$ or the in-place operation, $C=A \cdot B+C$.
\end{itemize}
where a \textit{fragment} is an overloaded class containing a tile of a matrix distributed across registers of all threads in the warp.

However, more operations are needed to execute FFT efficiently. (1) First, the data for FFT is usually stored in complex form. Therefore, in the stage of complex-matrix loading and storing, the input fragment of a complex matrix needs to be reformatted into an imaginary matrix fragment and a real matrix fragment. (2) Second, in the stage of computation, an element-wise multiplication operation is necessary to calculate $T_{N_1N_2} \odot X_{in}$ in equation \ref{equation:dft_mm}.

These two operations can only be done after the fragment is stored from registers to shared memory with WMMA APIs. But it is inefficient to do so, considering that eight Tensor Core units in one SM processor are capable of performing eight $4 \times 4 \times 4$ matrix-multiply-and-accumulate operation in one GPU clock cycle. 

% However, is not provided by NVIDIA, individual matrix elements must be accessed from memory (shared or global) after calling store\_matrix\_sync.

\medskip

\begin{figure}[h]
  \centering
  \includegraphics[width=\linewidth]{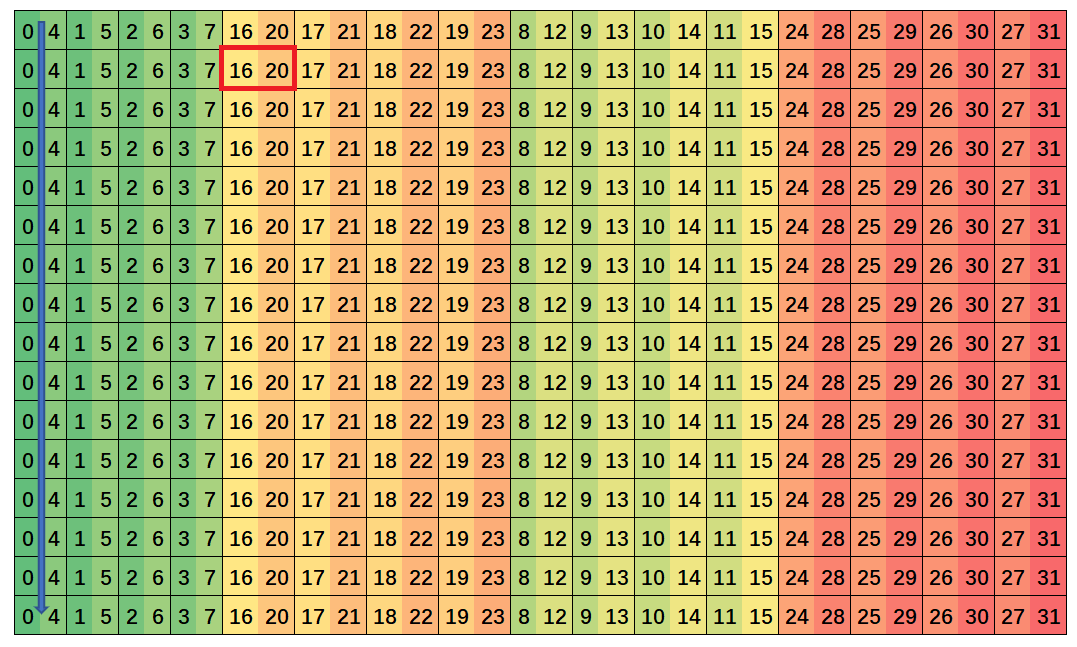}
  \caption{The map of the input fragment used in \fftlib{} on V100: The $16 \times 16$ matrix in the figure represents the matrix tile stored in this fragment. The numbers in the matrix indicate which threads in the warp store the element, for example, 16 and 20 in the second row and fifth column indicate that threads 16 and 20 have stored element $InFrag_{2, 5}$. The arrow in the first column shows the order of elements stored in thread 0, 4.}
  \label{fig:b}
\end{figure}

% In terms of the stage of data loading and storing, things are more complicated.
\noindent\textbf{A more flexible way to work with Tensor Cores.} NVIDIA does provide a method to access an individual fragment element but this can only be done uniformly, because we don't know which elements are stored in each thread. Prior work has showed some of these element distributed maps on specific GPU models. However, these maps differ when fragment parameters change and differ on different GPU models. So we developed a tool to help obtain specific maps when needed. The map of a \textit{half datatype, $16 \times 16 \times 16$ shape, row-major layout, matrix\_b type} fragment on V100 is shown in figure \ref{fig:b}. It is used to store tiles from the input data matrix in \fftlib{}. The $16 \times 16$ matrix in the figure represents the matrix tile stored in this fragment. The numbers in the matrix indicate which threads in the warp store the element, for example, 16 and 20 in the second row and fifth column indicate that threads 16 and 20 have stored the element $InFrag_{2, 5}$. The arrow in the first column shows the order of elements stored in thread 0, 4.

With the above maps, specified individual fragment element can be accessed. This individual element access is implemented using the fragment class member \textit{fragment::num\_elements} and the class member \textit{fragment::x[num\_elements]}. The first one keeps the num of fragment elements in this thread and the second one stores them in the above order.

\medskip

\noindent\textbf{Implementing FFT's special operations efficiently.}
% The special case of element-wise multiplication operation can be performed without use of shared memory, where threads in a warp will multiply all elements in the fragment by a same number. In this condition, direct element access can be implemented using the fragment class member \textit{fragment::num\_elements} and the class member \textit{fragment::x[num\_elements]}. The first one keeps the num of fragment elements in this thread and the second one is used to access them. With these fragment class members, uniform element-wise multiplication operation can be done within the registers instead of resorting to shared memory, with the following code:
% \begin{lstlisting}
% for (int i = 0; i < 
%      frag_real.num_elements; i++)
%   frag_real.x[i] = Const * frag_real.x[i];
% \end{lstlisting}
% This can be used to obtain $-A_{imag} \cdot B_{imag}$ from $A_{imag} \cdot B_{imag}$.
% We extend this method to non-uniform element-wise multiplication, with the above pattern of fragment. 
With the above method, we implemented efficient element-wise multiplications and complex matrix accesses to accelerate \fftlib{}. Moreover, we interleaved these two operations to hide latency.

Algorithm \ref{alg:fft_special} demonstrates this process. The \textit{calc\_eid} function takes the fragment element map and the thread ID in a warp as parameters and returns the element id in the matrix tile. Then the thread can load the element from memory while calculating the corresponding twiddle factor and performs a complex multiplication after that. Finally, the result is assigned to the Tensor Core fragments. This process replaces the original complex matrix decomposition and element-wise multiplication operations performed in shared memory.

% need to select the proper twiddle factors according to the elements it holds.

% In terms of the operation of element-wise multiplying with twiddle factors, it is a common case of element-wise multiplication operation. In this case, in order to use the above code to perform element-wise multiplication within registers, different threads in a warp need to select the proper twiddle factors according to the elements it holds. The code snippet demonstrates this process. Threads find their corresponding twiddle factors according to their thread IDs in the warp and the mapping pattern obtained by our tool, and then, perform element-wise production within registers. In addition, it is worth noting that by selecting the optimal type of \textit{matrix\_b}, the twiddle factors in a column can be recursively obtained by multiplication instead of calculating from scratch.

% \begin{figure}[h]
%   \centering
%   \includegraphics[width=\linewidth]{pics/code_snippet}
%   \caption{Code snippet: Threads find their corresponding twiddle factors according to their thread IDs in the warp and the mapping pattern, and then, perform element-wise production within registers.}
%   \label{fig:code_snippet_0}
% \end{figure}

\begin{algorithm}
  \caption{FFT's special operations}
  \label{alg:fft_special}
  \SetAlgoLined
  \LinesNumbered
  \tcc{Complex-matrix access and element-wise multiplication in a thread view}
  \KwIn{fragment element \textbf{map}}
  fragment frag\_in\_real\;
  fragment frag\_in\_imag\;
  \For{$i \leftarrow 0$ \KwTo $frag\_in\_real.num\_elements-1$}{
    eid $\leftarrow$ calc\_eid (threadId, i, \textbf{map})\;
    half2 twiddle $\leftarrow$ get\_twi(block\_start, warp\_start, eid)\;
    half2 in\_ele $\leftarrow$ In[block\_start+warp\_start+eid]\;
    in\_ele = cMul(in\_ele, twiddle)\;
    frag\_in\_real.x[i] $\leftarrow$ ele.x\;
    frag\_in\_imag.x[i] $\leftarrow$ ele.y\;
  }
\end{algorithm}

% When loading and storing data in complex form, we also use the above pattern.
% We use two fragments to load data from global memory. Each fragment has half of the complex-matrix tile, and then data are exchanged between the two fragments. Thanks to the type of \textit{matrix\_b}, the two fragments contain elements from the same column, so the exchanging of data only occurs within a thread. We also use two fragments to store the calculation results. We restore them into the continuous complex form before storing to the next level memory structure.

\medskip

By developing a tool to help find the map of matrix elements into each thread's fragment, we extended the programming methods of Tensor Cores, gained individual fragment element control, and showed how to use this method to optimize the \fftlib{} library. The effect of this optimization is shown and discussed in section \ref{sec:results}.

\subsection{Alleviate the Memory Bottleneck}
\label{sec:opt_mem}
Memory can easily become the bottleneck of FFT algorithms on GPU for two reasons: First, the original $\log_{}{N}$ merging processes of FFT algorithm require $\log_{}{N}$ times of global memory accesses, and the arithmetic intensity of a single merging is small. Second, merging processes in FFT require strided memory accesses, and uncoalesced strided accesses are quite inefficient on GPU. For the first problem, we have combined multiple mergings when designing \fftlib{}. This reduces the times of global memory accesses and increases the arithmetic intensity. For the second problem, we redesigned the data layout and memory access pattern as follows.

\medskip

\begin{figure}[tb]
  \centering
  \subfigure[Original out-of-place merging with a fixed data order]
  {
    \label{fig:cont_pattern_a}
    \includegraphics[width=0.9\linewidth]{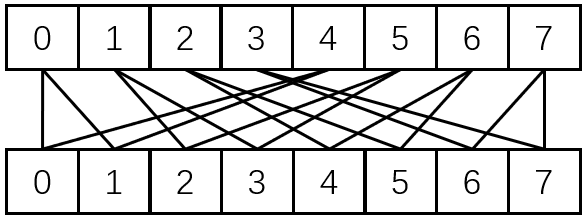}
  }
  \subfigure[\fftlib{}'s in-place merging with a changing data order; two adjacent butterflies are joined and warps can access memory with continuous size 2.]
  {
    \label{fig:cont_pattern_b}
    \includegraphics[width=0.9\linewidth]{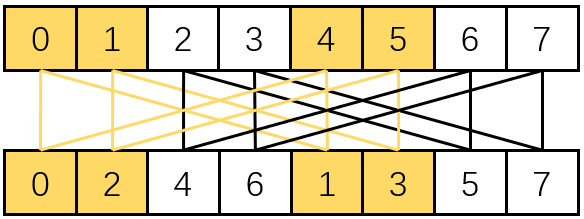}
  }
  \caption{\fftlib{}'s in-place and coalesced memory access pattern: the last radix-2 merging process of an 8-point sequence as an example.}
\end{figure}

\noindent\textbf{In-place computation data layout.} We first used an in-place computation data layout before coalescing memory accesses. Merging is an out-of-place process when data is stored in a fixed order in multiple iterations. Figure \ref{fig:cont_pattern_a} shows the last radix-2 merging process of an 8-point sequence as an example. The upper is the output, and the lower is the input. A pair of output elements is stored with a stride length 4. When implemented on GPU, an out-of-place algorithm requires twice the size of shared memory. This limits the continuous size that can be achieved by coalescing.

% Global memory accesses are serviced as memory transactions in GPU and a cache line is 128 bytes and maps to a 128 byte aligned segment in device memory. Memory accesses that are cached in both L1 and L2 are serviced with 128-byte memory transactions, whereas memory accesses that are cached in L2 only are serviced with 32-byte memory transactions. Transactions are issued in the granularity of a warp \cite{cudac++programmingguide}. This feature makes it important for us to pay attention to the memory access continuity of each warp when designing \fftlib{}, in order not to be slowned down by memory access.

% In a merging process, the sub-sequences are obtained by decimating according to a certain base number. This disrupts the continuity of the data. The previous algorithm transposes the output matrix to match the next input order. However, this part will consume a significant amount of time and slow down the speed of the entire FFT algorithm.

% In order to solve this problem, we combine multiple matrix transpositions into one data rearrangement, considering FFT is a linear transformation. The data rearrangement process take place before the data is transfered from CPU to GPU, and it is possible to realize an intermittent continuous memory access mode in all the next merge steps, without the need of matrix transpositions.

\fftlib{} stores data in a changing order in multiple iterations, figure \ref{fig:cont_pattern_b} shows the same merging process in \fftlib{}. It rearranges the original sequence according to parity. This allows the process to be executed in place. This rearrangement is executed recursively in multiple iterations to ensure that all mergings are in-place. The actual rearrangements are based on larger changing radices, although the principle is the same.

\medskip

\begin{table}[]
  \centering
  \caption{Achievable Global Memory Bandwidth under Different Continuous Size}
  \label{table:gbw}
  \begin{tabular}{cccc}
  \hline
  Cont. Sizes & Cont. Bytes & Mem. TP. (GB/s) & BLKs \\ \hline
  4 & 16 & 208.09 & 8 \\
  8 & 32 & 384.58 & 8 \\
  16 & 64 & 553.48 & 6 \\
  32 & 128 & 836.25 & 3 \\
  64 & 256 & 715.83 & 1 \\ \hline
  \end{tabular}
\end{table}

\noindent\textbf{Coalesced global memory accesses.} After in-place computation is implemented, a merging process includes multiple butterflies as shown in figure \ref{fig:cont_pattern_b}. Memory access in each butterfly is strided, but we can join adjacent butterflies together. In the figure, two adjacent butterflies are joined and warps can access memory with continuous size 2. Actually, in \fftlib{}, different continuous sizes are used for different radices.

% The native FFT memory access pattern is discrete, take the final radix-2 merging process of an 16-point sequence as an example, the memory access pattern in a butterfly is shown in picture \ref{fig:cont_pattern_a}. The data are discrete within a butterfly. However, we can join adjacent butterflies together. In picture \ref{fig:cont_pattern_b}, we let a warp to access multiple continuous data by join adjacent four butterflies.

Increasing the continuous size will increase continuity in memory access, but a bigger size makes a kernel use more shared memory and results in fewer concurrent blocks. To achieve higher performance, a proper continuous size is necessary. For radix-256 merging process as an example, the achievable global memory throughputs under different continuous sizes are shown in table \ref{table:gbw}. From the table, we can find that the achievable memory throughput increases as the continuous size increases when it is no more than 32. It is reasonable for the largest cache line size on GPU is 128 bytes. After that, the bandwidth drops instead. This is because that when the size exceeds 32, the number of concurrent blocks on a streaming multiprocessor reduces to one, and this will make the latency generated by block synchronization unable to be hidden.

\section{Evaluation}
\label{sec:results}
% In this section, we evaluate our \fftlib{} in comparison with cuFFT, the state-of-the-art FFT library on GPU.

\subsection{Experimental Setup}

\noindent\textbf{Methods.} We compare \fftlib{} with NVIDIA cuFFT which is the state-of-the-art FFT library on GPU. Other GPU FFT libraries are either not open source or slower than cuFFT. The version of cuFFT we used is 11.0 released in Aug. 2020. As of the time of this writing, it is the latest version on our testing DGX-2 platform and DGX-A100 platform. We compare them in terms of accuracy and performance. We show the performance of batched 1D and 2D FFTs of adequate sizes on Tesla V100 GPU and Tesla A100 GPU to evaluate the generalization of our algorithm. 

\medskip

\noindent\textbf{Platform} We measured the performance of \fftlib{} on two platforms, as shown in Table \ref{table:platform}, including two generations of NVIDIA GPUs (Volta, and Ampere microarchitectures). 

\begin{table}[H]
  \centering
  \caption{Platform Information.}
  \label{table:platform}
  \begin{tabular}{ccc}
  \hline
  Platform & Volta                    & Ampere        \\ \hline
  GPU      & Tesla V100               & Tesla A100    \\
  CPU      & Intel Xeon 8168 & AMD Rome 7742 \\
  OS       & CentOS 7                 & CentOS 7      \\ 
  CUDA     & 11.0                     & 11.0          \\ \hline\hline
  \begin{tabular}[c]{@{}c@{}}Peak FP16\\ (CUDA Core)\end{tabular} & 31.4 TFlops   & 78 TFlops          \\ \hline
  \begin{tabular}[c]{@{}c@{}}Peak FP16\\ (Tensor Core)\end{tabular} & 125 TFlops       & 312 TFlops          \\ \hline
  Memory Bandwidth     & 900 GB/sec   & 1555 GB/sec   \\ \hline
  \end{tabular}
\end{table}

\medskip

\noindent\textbf{TestCase.} In the 1D performance test, we measured batched 1D FFTs of short length, moderate length and long length, from 256 to 134,217,728. For each length, we used a batch size big enough to fully utilize all the Streaming Multiprocessors and Tensor Cores. As for 2D FFT, we measured the six common lengths with adequate batch size. In the follow-up batch size test, we fixed the length, and then measured the performance versus different batch sizes. For all the tests, the input sequences were generated randomly in the interval -1.0 to 1.0.

\medskip

\noindent\textbf{Performance Metric.} In the performance tests, the data are first transferred from CPU to GPU and a plan is created. Then, the \textit{execute function} are executed thousands of times and the average performance is reported. The time spent on the data transferring and plan creating are not counted, for a plan can be reused during the whole life of real applications. We use radix-2 equivalent Trillion Floating Point Operations per Second (TFLOPS) as the Performance metric, because the total number of calculations depends on the specific radix. It can be calculated with equation \ref{equation:perf_def}.
\begin{equation}
  \small
  \label{equation:perf_def}
  TFLOPS = \frac{6 \times 2 \times \log_{2}{N} \times N \times N\_Batch \times RepeatingTimes}{TotalTime \times 10^{12}} 
\end{equation}
where N denotes the length of a sequence and N\_Batch denotes the num of sequences. 

\medskip

\noindent\textbf{Precision Metric.} \fftlib{} uses different FFT radices in the divide-and-conquer progress from common FFT algorithms, which influences the result to some degree, so we compare the relative error which is defined as equation \ref{equation:prec_def}.
\begin{equation}
  \small
  \label{equation:prec_def}
  RelativeError(X) = \frac{1}{N} \sum_{i=0}^{N} \left | \frac{X_{double}[i] - X[i]}{x_{double}} \right |
\end{equation}
where $X_{double}$ denotes the sequence calculated by the FFTW library in double precision. It is used as the standard result.

\subsection{Precision}

% We compare our \fftlib{} and the vendor FFT library cuFFT using three kinds of test cases: different length 1D FFT of adequate batch sizes, different length 2D FFT of adequate batch size and different batch size 1D/2D FFT.

Table \ref{table:avg_error} shows the average relative error comparison between our \fftlib{} and cuFFT in 1D and 2D cases.

\begin{table}[htbp]
  \small
  \centering
  \caption{Average relative error of 1D and 2D FFT, comparison between cuFFT and \fftlib{}}
  \label{table:avg_error}
  \begin{tabular}{lllll}
    \hline
    & cuFFT-1D & \fftlib{}-1D & cuFFT-2D & \fftlib{}-2D\\\hline
    Relative Error & 1.78$\pm$0.5\% & 1.76$\pm$0.5\% & 1.65$\pm$0.1\% & 1.65$\pm$0.1\%\\\hline
  \end{tabular}
\end{table}

\fftlib{} uses different merging radices from cuFFT, and the calculation processes of the two libraries differ. However, from the table, we can find that comparing with the standard result calculated by double precision FFTW, the error of the two libraries is at the same level. Although in theory, \fftlib{} uses matrix multiplication as the basic operator, which has better error control. But the complete progress of FFT consists of multiple mergings. It is equivalent to performing multiple matrix multiplications in turn and storing intermediate results in half precision. The storage of intermediate results is the main source of error. This process is similar in our \fftlib{} and cuFFT.

\subsection{Overall Speedup}

\begin{figure*}[htb]
  \centering
  \subfigcapskip=-5pt
  \subfigure[on V100]
  {
    \label{fig:res_1d_perf_V100}
    \includegraphics[width=0.45\linewidth]{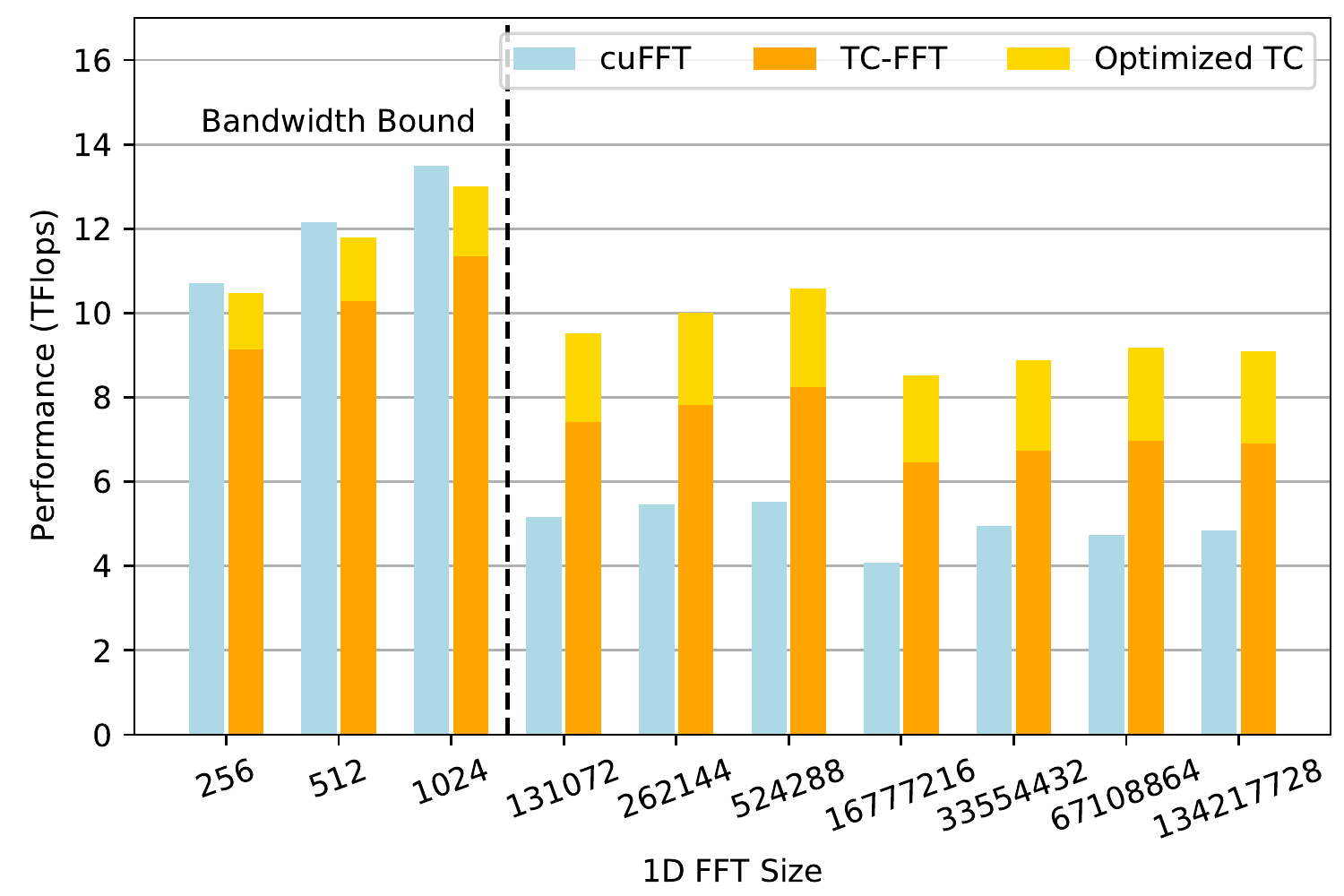}
  }
  % \quad
  \subfigure[on A100]
  {
    \label{fig:res_1d_perf_A100}
    \includegraphics[width=0.45\linewidth]{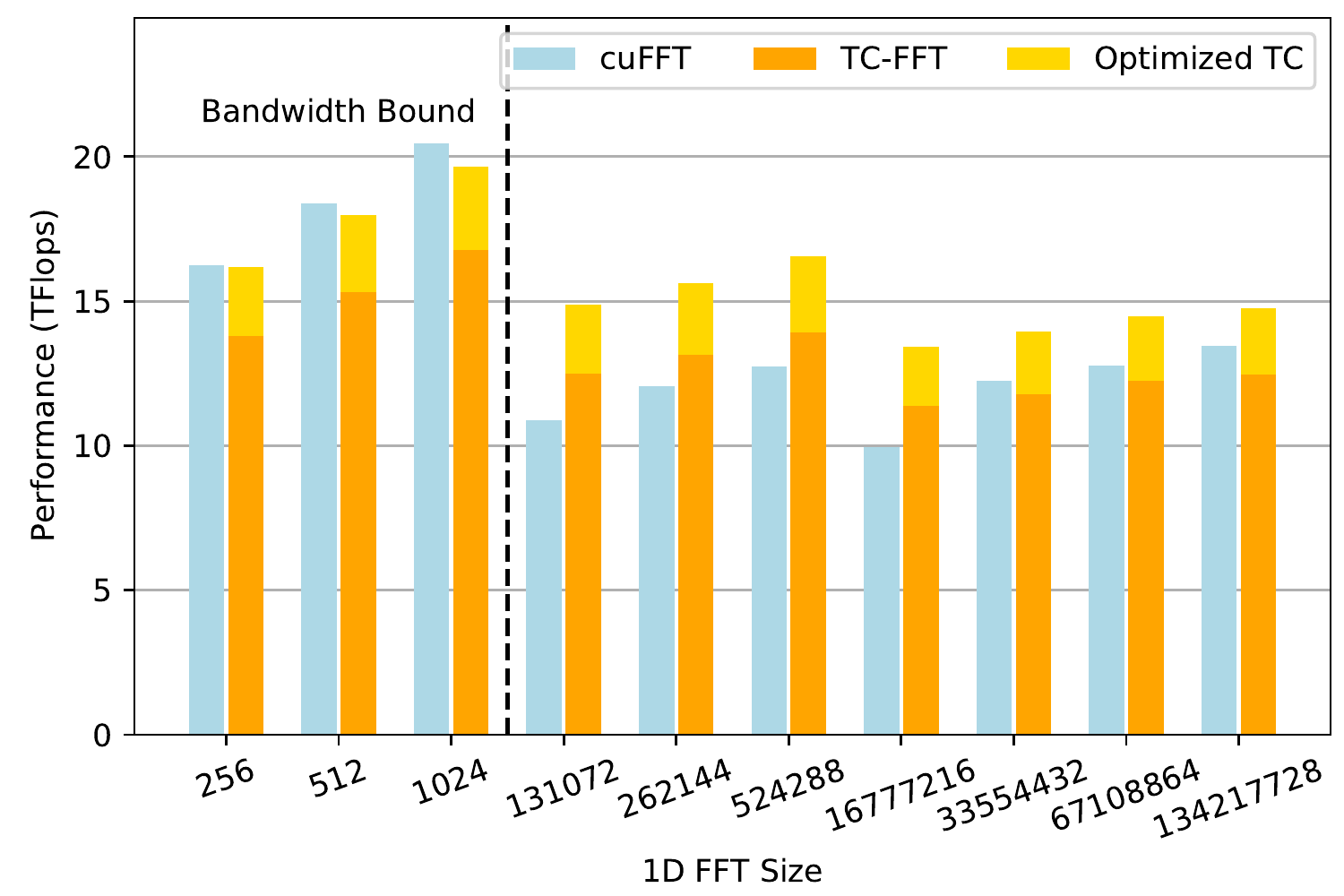}
  }
  \vspace{-1\baselineskip}
  \caption{Performance of 1D FFT of different sizes on two platforms.}
  \label{fig:res_1d_perf}
\end{figure*}

\begin{figure*}[htb]
  \centering
  \subfigcapskip=-15pt
  \subfigure[on V100]
  {
    \label{fig:res_2d_perf_V100}
    \includegraphics[width=0.45\linewidth]{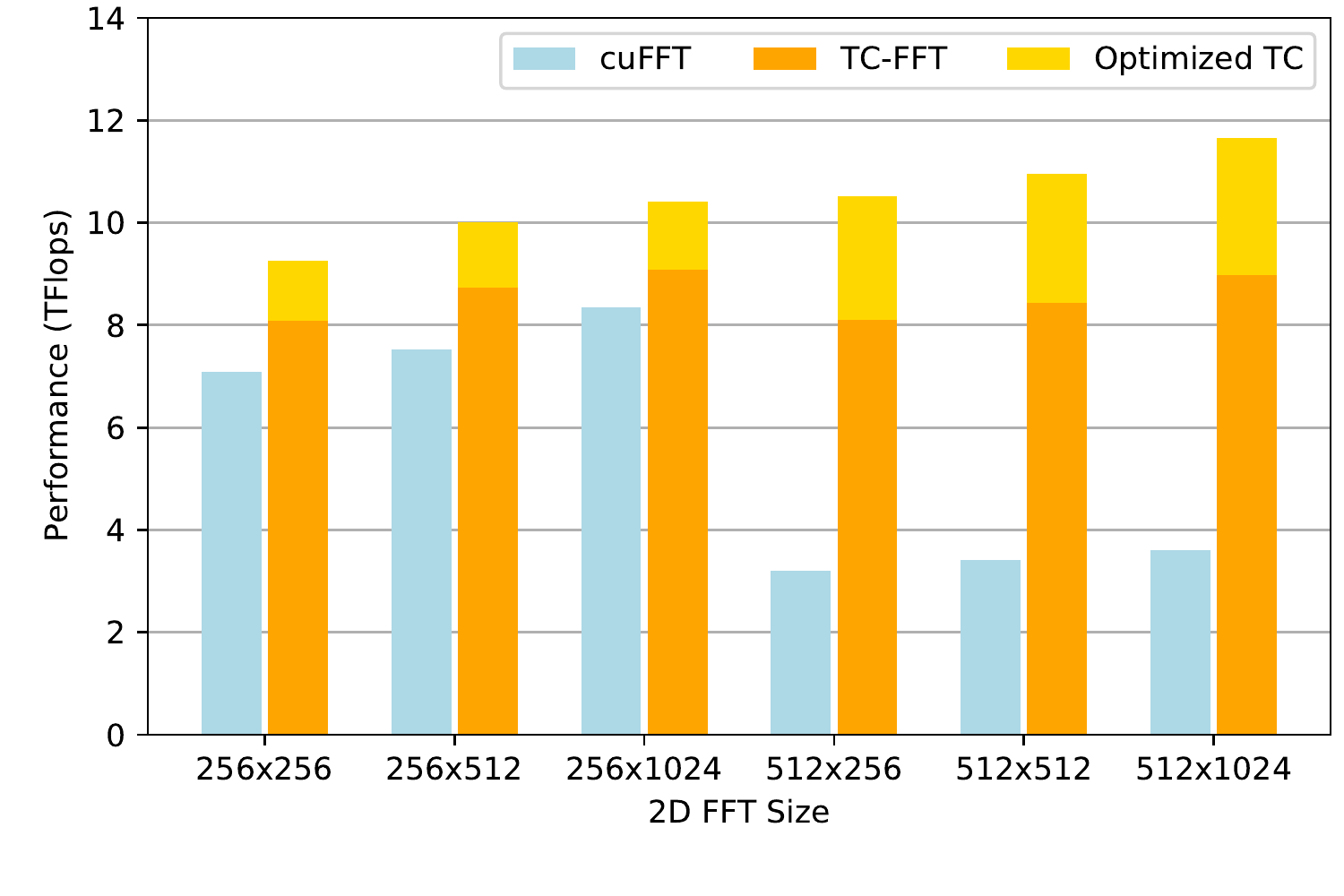}
  }
  \subfigure[on A100]
  {
    \label{fig:res_2d_perf_A100}
    \includegraphics[width=0.45\linewidth]{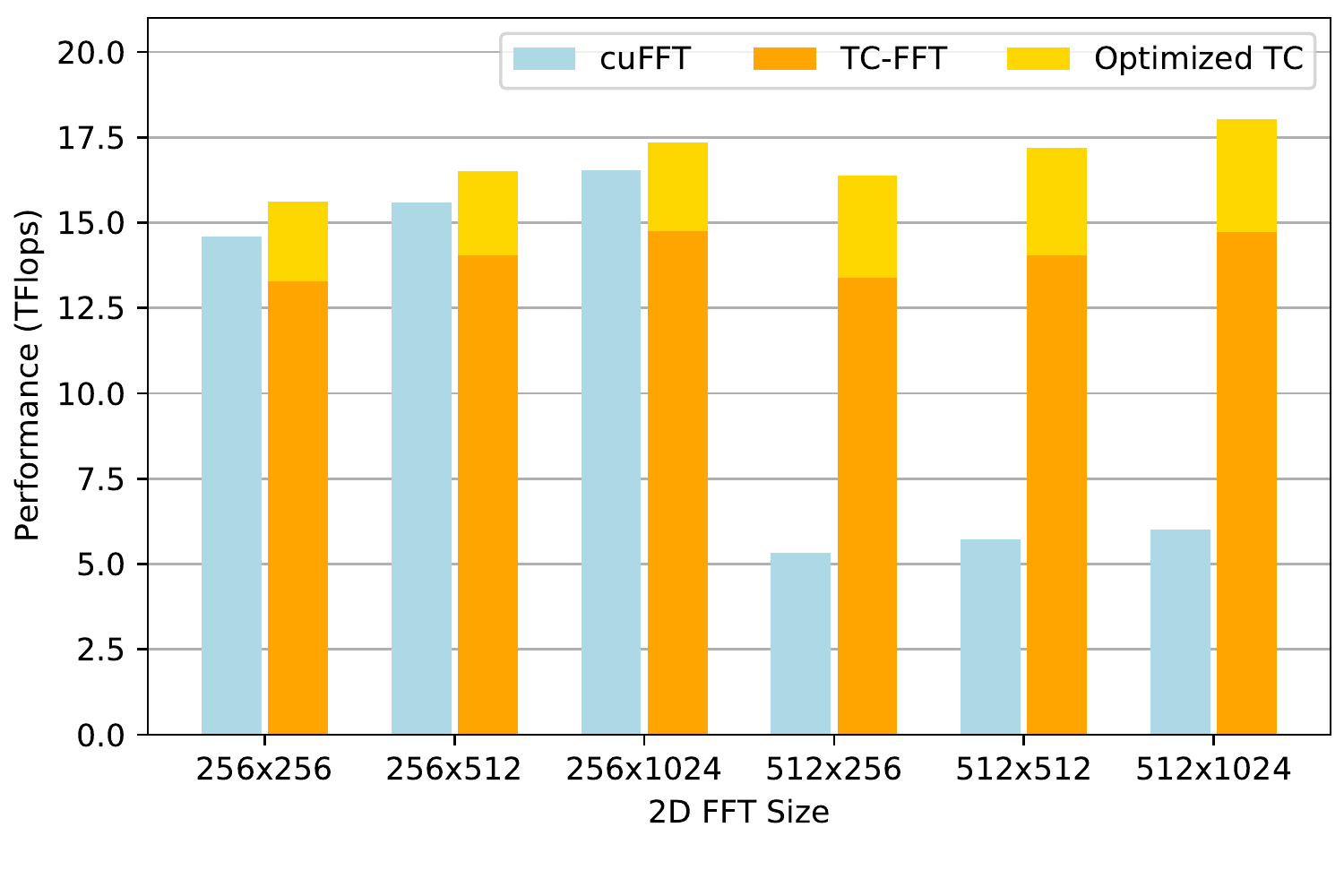}
  }
  \vspace{-1\baselineskip}
  \caption{Performance of 2D FFT of different sizes on two platforms.}
  \label{fig:res_2d_perf}
\end{figure*}

\begin{figure*}[htb]
  \centering
  \subfigcapskip=-5pt
  \subfigure[1D FFT]
  {
    \label{fig:res_1d_mem_V100}
    \includegraphics[width=0.45\linewidth]{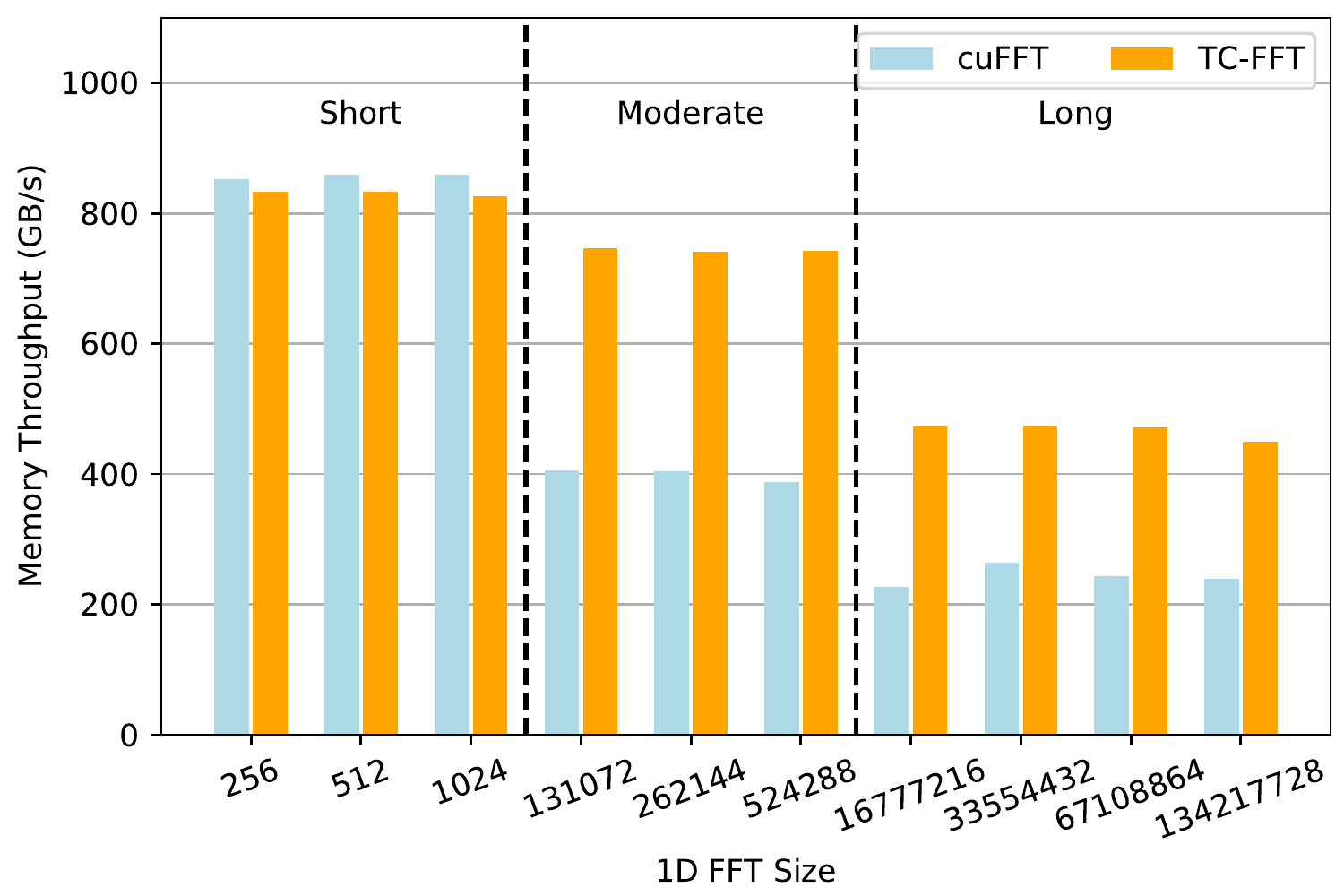}
  }
  \subfigure[2D FFT]
  {
    \label{fig:res_2d_mem_V100}
    \includegraphics[width=0.45\linewidth]{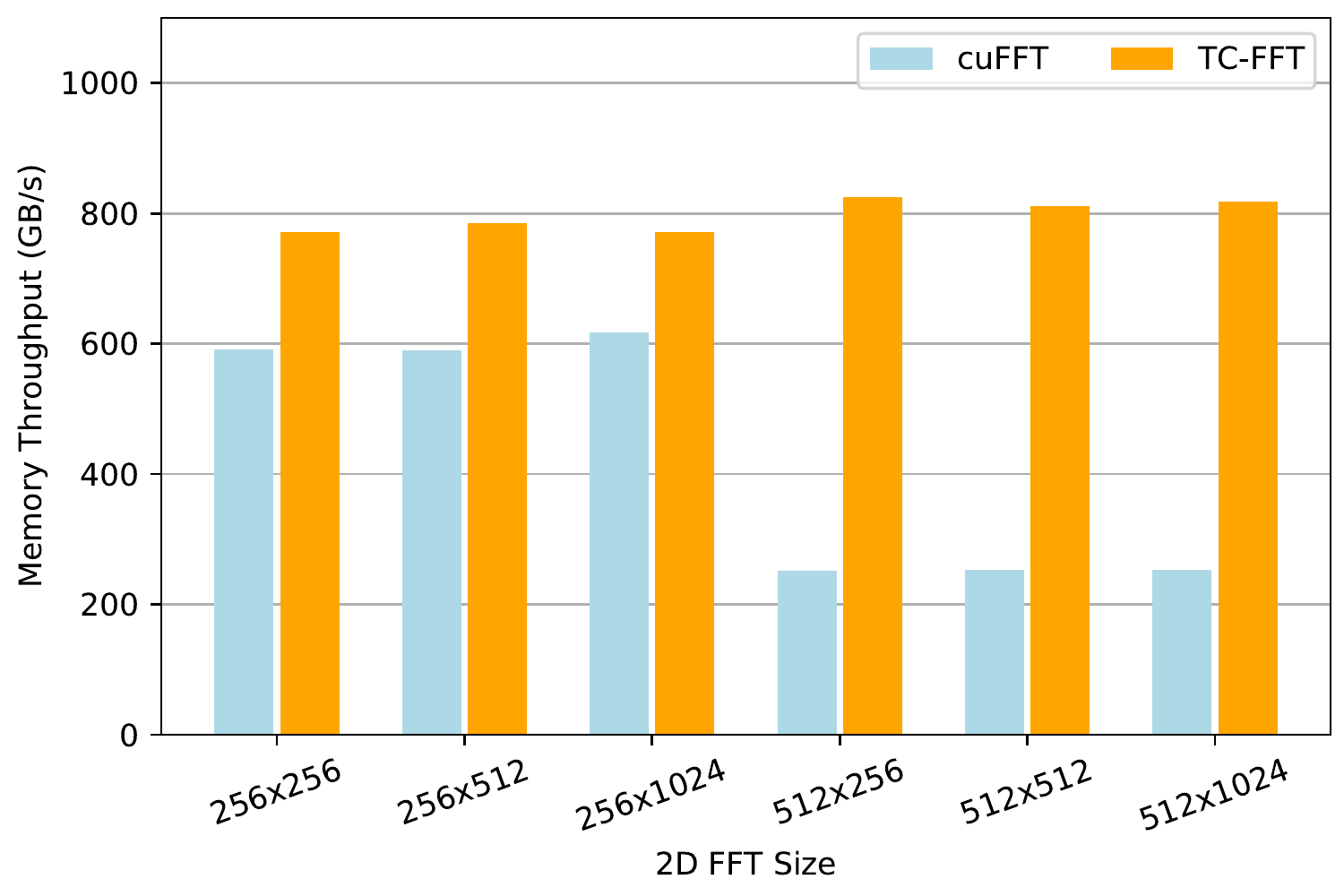}
  }
  \vspace{-1\baselineskip}
  \caption{Global memory bandwidth of 1D and 2D FFT of different sizes on V100.}
  \label{}
\end{figure*}

\medskip

\noindent\textbf{1D FFT.} Figure \ref{fig:res_1d_perf_V100} shows the performance comparison between our \fftlib{} and cuFFT on V100. As shown in the figure, the result can be roughly divided into two parts, bandwidth-bound cases and other cases. In the bandwidth-bound cases, a single sequence is short enough to be completely put into the shared memory and the core merging kernels is simple with no need of block-range thread synchronizations. Calculations can be totally overlap with memory accesses and the performance is mainly limited by the global memory bandwidth. In these cases, the memory throughput of cuFFT is close to the theoretical bandwidth peak, and our \fftlib{} can reach 96.4\% to 97.8\% performance of cuFFT. When the FFT size grows, the core merging kernels begin to use thread synchronizations which makes some calculations no longer overlap with memory accesses. In these cases, the time spent on the calculation begins to affect the overall performance and our \fftlib{} can achieve a minimum 1.84x speedup and an average 1.90x speedup compared with cuFFT. 
Figure \ref{fig:res_1d_perf_A100} shows the results on A100. The results has the similar pattern to those on V100. In the bandwidth-bound cases, our \fftlib{} can achieve 96.1\% to 99.7\% performance of cuFFT. In other cases, it can achieve an average of 1.24x speedup. The benefits on A100 are less than those on V100. One reason is that, compared to V100, A100 has 2.5x half-precision computing power but only a 1.7x global memory bandwidth. As a result, optimized FFT algorithm that resorts to more computing power can only bring less performance gain.

\medskip

\noindent\textbf{2D FFT.} Figure \ref{fig:res_2d_perf_V100} shows the performance of doing FFT of batched 2D sequences on V100. 2D FFT performs 1D FFTs in turn on each dimension of the data. Like cuFFT and FFTW, \fftlib{} use row-major order to store 2D sequences. This means that the data in the first dimension does not continue in memory and the FFT along this dimension is the main performance factor. When the size of the first dimension is 256, \fftlib{} is faster than cuFFT by 1.29x on average. And when the first dimension is 512 \fftlib{} is faster by 3.24x on average. Like long length 1D FFT, mergings along the first dimension require thread synchronizations, so this benefit comes from both improvement in calculation speed and our data arrangement design. Figure \ref{fig:res_2d_perf_A100} shows the results on A100, similar to the situation of 1D FFT, performance improvement brought by \fftlib{} on A100 is also less than than on V100. However, \fftlib{} is still faster than cuFFT by 3.03x when the first dimension is 512.

\medskip

The above results show that, in bandwidth-bound cases, \fftlib{} can use up almost all the bandwidth. And in non-bandwidth-bound cases, \fftlib{} can outperform cuFFT by a notable margin across different GPU architectures. This significant benefit comes from the extremely high computation power of Tensor Cores and our novel kernel optimizations.

\subsection{Analysis}
Since similar patterns are shown on V100 and A100, we will focus on the results on V100 in the following analysis.

\begin{figure*}[h]
  \centering
  \subfigcapskip=-10pt
  \subfigure[1D FFT: 131072]
  {
    \label{fig:res_b_131072}
    \includegraphics[width=0.45\linewidth]{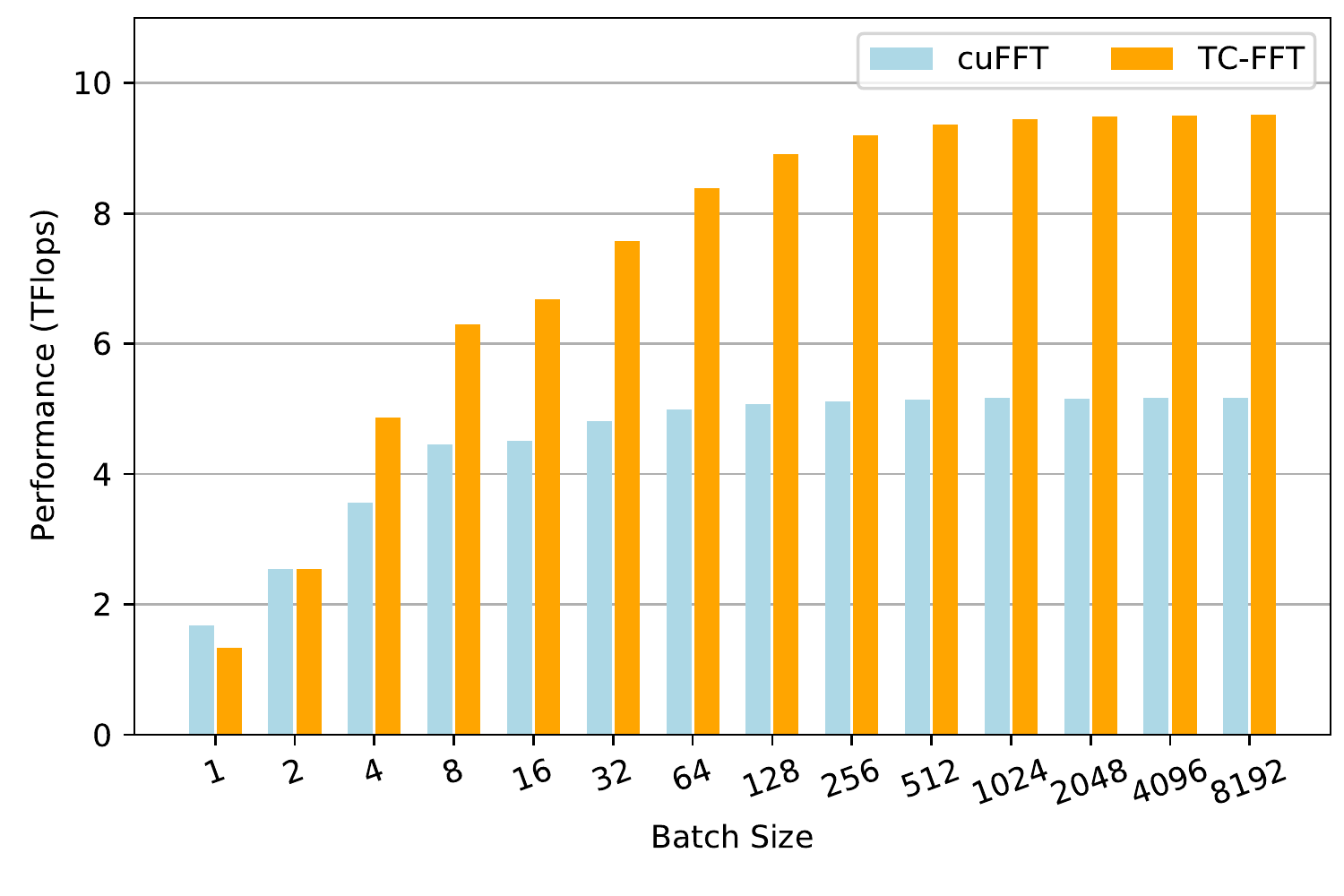}
  }
  \subfigure[2D FFT: $512 \times 256$]
  {
    \label{fig:res_b_512x256}
    \includegraphics[width=0.45\linewidth]{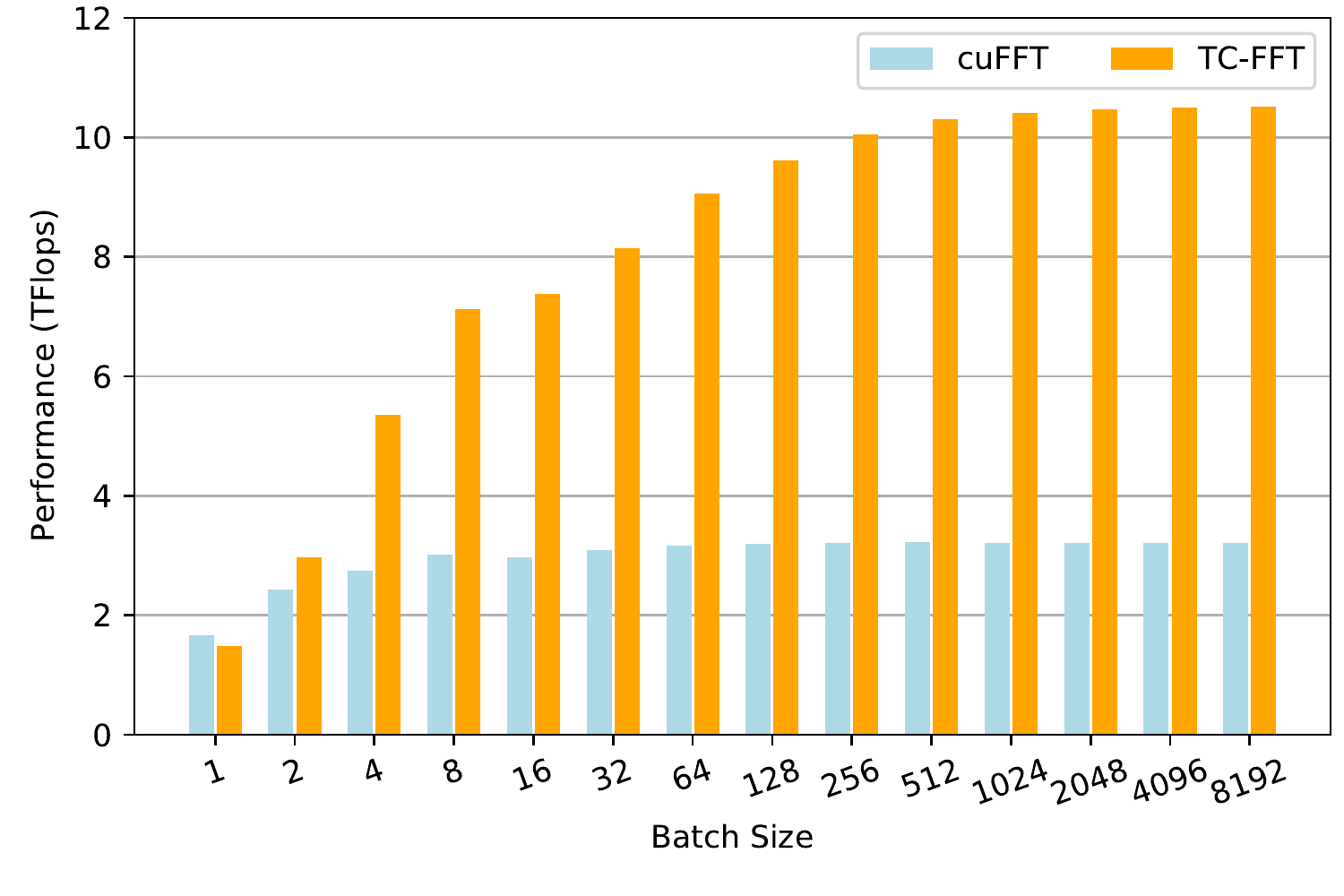}
  }
  \vspace{-1\baselineskip}
  \caption{Performance of 1D and 2D FFT of different batch sizes on V100.}
  \label{}
\end{figure*}

\medskip

\noindent\textbf{Benefit of optimizations to Tensor Cores.} We extended the programming methods of Tensor Cores with the fragment storage map, gained element-level control over Tensor Core fragments, and used this method to optimize \fftlib{}. With element-level control, \fftlib{} can accomplish element-wise multiplication and complex-matrix accesses without use of shared memory. This greatly reduces the latency of these operations. We shows the performance benefits of this optimization in both figure \ref{fig:res_1d_perf} and figure \ref{fig:res_2d_perf} with \textit{Optimized TC} label. From the figures, we can find this optimization brings 1.15x-1.32x speedup. 

\medskip

\noindent\textbf{Benefit of data arrangement redesign.} Figure \ref{fig:res_1d_mem_V100} shows the global memory throughput of different size 1D FFTs. In the figure, we divide the sizes of 1D FFTs into three parts: short, moderate, and long, according to the required stride lengths of memory accesses. FFTs of larger sizes require longer stride lengths and use merging kernels with less computation overlap, so their memory throughput is lower. In short cases, thanks to our redesign of data arrangement and the memory access pattern, the memory throughput of \fftlib{} is close to the peak global memory bandwidth, and in other cases, \fftlib{} can outperform cuFFT nearly 2x.

Things are similar in 2D FFT cases. Figure \ref{fig:res_2d_mem_V100} shows the global memory throughput of different size 2D FFTs. The elements in the first dimension of 2D FFTs are scattered in the memory. From the figure, we can find that \fftlib{} obviously exceeds cuFFT in all the cases and when the size of the first dimension increases the performance of cuFFT drops a lot while that of \fftlib{} almost remains the same.
%This proves that our method has good consistency on different merging radices.

\medskip

\noindent\textbf{Performance of small batch sizes.} The performance results above are measured with a batch size big enough to fully utilize the GPU resources. Here we give the performance results of smaller batch sizes. Figure \ref{fig:res_b_131072} shows the performance comparison between our \fftlib{} and cuFFT when transforming batched 1D 131072-point sequences. \fftlib{} is faster than cuFFT when batch size is larger than 4, and the speedup ratio gradually increases. Figure \ref{fig:res_b_512x256} shows the performance of batched 2D $512 \times 256$ FFTs. \fftlib{} begins to outperform cuFFT when batch size is 2. This result shows that \fftlib{} also performs well on small batch sizes.

\section{Related Work}

Accelerating HPC workload through AI-specific hardware, such as Tensor Cores, has attracted many research efforts. This work is broadly related to the researches under three topics: 

\medskip

\noindent\textbf{Utilizing Tensor Cores for dense linear algebra.} Some work utilized Tensor Cores to accelerate dense linear algebra in the performance critical steps \cite{abdelfattah2020survey, haidar2017investigating, yan2020demystifying, mukunoki2020dgemm}. Haidar \cite{haidar2018harnessing} utilized Tensor Cores to accelerate LU factorization which solves a system of equations. EGEMM-TC \cite{feng2021egemm} used precision recovery GEMM on Tensor Cores to accelerate some scientific applications. Most of the existing work focused on GEMM, convolution, or other dense linear algebra which meet the primitive of Tensor Cores. In contrast, \fftlib{} exploits Tensor Cores in FFT which is more challenging to be re-expressed with matrix-matrix operators.

\medskip

% \noindent\textbf{Tensor Core for HPC Applications.} Some research efforts have been devoted towards accelerating high performance computing workloads with Tensor Cores. Hatfield \cite{hatfield2019accelerating} accelerate the calculation of the Legendre transforms in the Integrated Forecasting System (IFS) with Tensor Core. Ichimura \cite{ichimura2018fast} utilize FP16 arithmetic to improve the convergence of iterative solver leading to 5.56-fold reduction in arithmetic count from a standard solver.

\medskip

\noindent\textbf{Exploiting Tensor Cores in FFT.} Sorna \cite{sorna2018optimizing} and Cheng \cite{cheng2018accelerating} gived the theoretical basis of using Tensor Cores in FFT and their example implementation resorted to cuBlas to utilized Tensor Cores. But the performance of their implementation was far inferior to cuFFT. Durran \cite{durrani2021fft} has proposed an optimized 1D FFT algorithm in their poster. Their implementation with Tensor Core WMMA APIs outperformed cuFFT and used shared memory to improved the arithmetic intensity, but only on the basic small size 1D FFT. They did not deal with the memory bottleneck caused by the unique memory access pattern of large size or multidimensional FFT, and there is still considerable room for improvement in their method to support FFT's special operations. Different from prior work, our experimental results show that \fftlib{} can achieve higher performance than cuFFT in 1D and 2D FFT of universal sizes.

\medskip

\noindent\textbf{Implementing ditributed FFT on Heterogeneous System.} Several approaches have been proposed for large size ditributed FFT on heterogeneous systems. Some work, like PFFT \cite{pippig2013pfft} and P3DFFT \cite{pekurovsky2012p3dfft}, focused on how to implement distributed FFT in the highly efficient and scalable way. heFFTe \cite{ayala2020heffte, ayala2019impacts} built a multi-node communication model for distributed FFT and achieved more than 2× speedup for the whole FFT computation. Their work focuses on the optimization of communication between computing nodes, and \fftlib{} can be integrated to accelerate the FFT on each node.

\section{Conclusion}

% The Fast Fourier Transform is a essential tool in scientific and engineering computation. And the demand for lower-precision or mixed-precision FFTs is gradually increasing, which make it possible to utilize half precision floating point arithmetic (FP16) to get faster speed and save energy. NVIDIA Tensor Cores achieve extremely high performance with half precision which designed for deep learning workloads. However, the fixed computation pattern and lower precision limit the application of Tensor Cores especially in the area of scientific computing. 

We design and implement \fftlib{} which utilizes Tensor Cores to accelerate half-precision FFT in both 1D and 2D forms of various sizes. And we exploit a set of optimizations to efficiently support FFT's special operations and to alleviate memory bottleneck. We evaluate \fftlib{} and the NVIDIA cuFFT in various sizes and dimensions on the latest two generations of NVIDIA GPUs, V100 and A100. The results show that our \fftlib{} can outperform cuFFT 1.29x-3.24x and 1.10x-3.03x on the two GPUs, respectively. \fftlib{} shows a great potential to use this AI-specific hardware to accelerate FFT and the methods of \fftlib{} can be generalized to higher precision on higher precision Matrix Operation Units.

We have identified three major avenues for future work. First, the current version of \fftlib{} achieves a significant performance improvement, while it only supports half-precision. Follow-up efforts can be made to support higher precision on higher precision Matrix Operation Units; Second, \fftlib{} has no consideration of precision recovery. We will try to introduce some precision recovery algorithms to improve the precision of \fftlib{} on low precision Matrix Operation Units; Finally, as \fftlib{} can provide a significant speedup, we will use \fftlib{} in some real-world scientific applications to further confirm that \fftlib{} can extend the usage of Tensor Core in scientific computing.

%%
%% The acknowledgments section is defined using the "acks" environment
%% (and NOT an unnumbered section). This ensures the proper
%% identification of the section in the article metadata, and the
%% consistent spelling of the heading.
%\begin{acks}
%We thank HPC Center of Shanghai Jiao Tong University for providing computing resource and excellent technical support.
%\end{acks}

%%
%% The next two lines define the bibliography style to be used, and
%% the bibliography file.
\bibliographystyle{ACM-Reference-Format}
\bibliography{main}

\end{document}